# Modularity affects the robustness of scale-free model and real-world social networks under betweenness and degree-based node attack

Q. Nguyen [a,b], T.V. Vu [a], H.-D. Dinh [c], D. Cassi [d], F. Scotognella [e], R. Alfieri [d], M. Bellingeri [d,e]

[a] Division of Computational Mathematics and Engineering, Institute for Computational Science, Ton Duc Thang University, Ho Chi Minh City, Vietnam

[b] Faculty of Finance and Banking, Ton Duc Thang University, Ho Chi Minh City, Vietnam

[c] John Von Neumann Institute, Vietnam National University Ho chi minh City, Ho chi minh City, Vietnam

[d] Dip. Scienze Matematiche, Fisiche e Informatiche, Università di Parma, Parco Area delle Scienze, 7/A, 43124 Parma

[e] Dipartimento di Fisica, Politecnico di Milano, Piazza Leonardo da Vinci 32, 20133, Milano, Italy

Corresponding author email address: nguyenquang@tdtu.edu.vn

## Abstract

In this paper we investigate how the modularity of model and real-world social networks affect their robustness and the efficacy of node attack (removal) strategies based on node degree (ID) and node betweenness (IB). We build Barabasi-Albert model networks with different modularity by a new *ad hoc* algorithm that rewire links forming networks with community structure. We traced the network robustness using the largest connected component (*LCC*). We find that higher level of modularity decreases the model network robustness under both attack strategies, i.e. model network with higher community structure showed faster *LCC* disruption when subjected to node removal. Very interesting, we find that when model networks showed non-modular structure or low modularity, the degree-based (ID) is more effective than the betweenness-based node attack strategy (IB). Conversely, in the case the model network present higher modularity, the IB strategies becomes clearly the most effective to fragment the *LCC*. Last, we investigated how the modularity of the network structure evaluated by the modularity indicator (*Q*) affect the robustness and the efficacy of the attack strategies in 12 real-world social networks. We found that the modularity *Q* is negatively correlated with the robustness of the real-world social networks under IB node attack strategy ($p$-value$< 0.001$). This result indicates how real-world networks with higher modularity (i.e. with higher community structure) may be more fragile to betwenness-based node attack. The results presented in this paper unveil the role of modularity and community structure for the robustness of networks and may be useful to select the best node attack strategies in network.

**Introduction**

The study of real-world complex networks has attracted much attention in recent decades because a large number of complex systems in the real world can be considered as complex networks, such as social (Borgatti et al. 2009, Bellingeri et al. 2020), technological (Albert et al. 1999, Faloutsos et al. 1999), biological (Jeong et al. 2000, Barra et al. 2010), ecological complex systems (Bellingeri and Bodini 2013; Bellingeri and Vincenzi 2013). Many real-world networks show a scale-free structure, making them resilient to random node failure (Cohen et al. 2000) but can disintegrate quickly when a small proportion of important nodes are removed (Albert et al. 1999). The network's robustness, which evaluates the capability of network to hold its functioning under such failures or attacks has drawn extensive attention in recent years (Albert and Barabási 2002; Cohen et al. 2000; Callaway et al. 2000; Iyer et al. 2013; Bellingeri et al. 2015; Bellingeri et al. 2014; Dall'Asta et al. 2006; Nguyen and Nguyen 2018; Wandelt et al. 2018; Bellingeri et al. 2019;2020). Usually, Monte-Carlo simulation is used to evaluate the network robustness: for random failure, nodes/edges are removed with the same probability (random removal), while for intentional attack, nodes/edges are removed according to different structural properties of the network and a robustness measure is then computed during the node/edge removal simulation (Albert et al. 2000; Cohen et al. 2000, 2001; Bellingeri et al. 2020; Lekha and Balakrishnan 2020). To identify the node/edge removal strategy that triggers the greatest amount of damage in the system is also highly important for revealing the links/nodes that act as key players in network functioning with many practical applications (Bellingeri et al. 2020). For example, the understanding of how the node/edge removal affects real social systems may predict how the abandoning of individuals affects the information spread in the social network, thus individuating the "influential spreaders" in the network, such as most important scholars or influencers (Ahajjam and Badir 2018; Bellingeri et al. 2020). On the other hand, in social contact network on which a disease can spread, it is critical to understand how node removal through vaccination affects the spread of the disease to efficiently prevent an epidemic (Holme 2004; Wang et al. 2015; Bellingeri et al. 2020).

One of the most important measure of network robustness is the size of the largest connected component (*LCC*), i.e. the *LCC* is the highest number of connected nodes in the network (Albert et al. 2000). The *LCC* gives us a simple interpretation of the system robustness when subjected to node/edge removal accounting the largest functioning part of the network. For example, if the Internet is attacked, all nodes (servers) within the *LCC* can still transfer information mutually and indicating the largest networked structure still active. Another example, in a social contact network, the *LCC* represents the highest number of individuals that can be affected by a disease spreading (Bellingeri et al. 2019). For this reason, the most efficient node attack strategy is the one that is able to induce the fastest *LCC* decrease (Figure 1), and numerical simulations have shown that attack strategies based on network's nodes centrality measures can effectively individuating the most important nodes to reduce the size of the *LCC* (Albert et al. 2000, Cohen et al. 2000, 2001, Callaway et al. 2000; Iyer et al. 2013; Bellingeri et al 2018; Bellingeri et al. 2014; Nguyen and Nguyen 2018; Wandelt et al. 2018). In specific, overall findings showed that nodes attack strategies based on betweenness centrality are highly efficient to dismantle the *LCC* (Iyer et al. 2013, Bellingeri et al. 2014; Sun et al. 2017, Nguyen and Nguyen 2018; Wandelt et al. 2018), especially for real-world networks. However, the difference in the effectiveness varied considerably among different real-world networks (Iyer et al. 2013; Bellingeri et al. 2014; Wandelt et al. 2020).

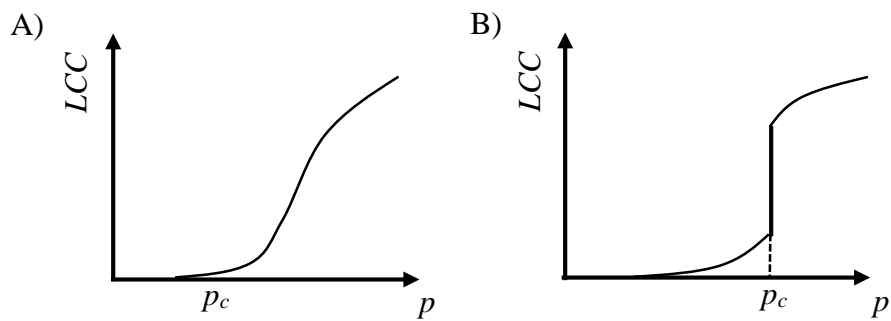

**Figure 1**: Schematic behavior of the size of the *LCC* as a function of the proportion of remaining nodes $p$ during a node removal process: A) *LCC* shows a continuous $2^{nd}$ order decreases without abrupt decrease and B) *LCC* is subjected to first-order percolation phase transition (Achilioptas et al. 2009) showing an abrupt decrease in correspondence of $p=p_c$. The node attack strategy in B is able to dismantle

the network with a small proportion of removed nodes, i.e. is the most effective to decrease the *LCC*, thus individuating the most important nodes in network.

The mechanism that gives rise to such an abrupt decrease is studied using percolation theory and is assigned to the first-order percolation phase transition (Achilioptas et al. 2009, Riordan et al. 2011, Cho et al. 2013). However, the question whether such an abrupt decrease occurs for a certain real network under attack remain unclear. This question is of great importance from two aspects: on one hand, if we want to break a network using node removal, we would find strategies that remove nodes that can cause such abrupt and fast decrease in *LCC*'s size. On the other hand, if we want to protect a network, we must design it in a way that such abrupt decrease should not happen. Since the network robustness must depend on its topology, several studies have investigated the relationship between topological metrics and the robustness of a network.

Iyer et al. (2013) studied robustness of model networks with power-law and exponential degree distribution, with various node clustering coefficient (or node transitivity) level. They found that increasing the clustering coefficient of the network nodes results in decreasing robustness to node attack with the most dramatic effect being displayed for node attack based on their degree and betweenness. The authors also suggested for increasing the robustness, it is necessary to design topological structures with low clustering coefficient as is consistent with the functional requirements of the network. Their simulation on real-world networks also show that the difference in the effectiveness between strategies varied across networks.

Nguyen and Trang (2019) studied the Facebook social networks and found those networks with higher modularity *Q* have lower robustness to node removal. The modularity indicator *Q* introduced by Newman and Girvan (2004) measures how well a network breaks into communities, (i.e. a community or module in a network is a well-connected group of nodes which have sparser connections with the nodes outside the group). Networks with high modularity *Q* have dense connections (more links) among the nodes within modules but sparse connections (few links) among nodes from different modules. Therefore, the modularity *Q* is higher in networks with marked community structure, which are called modular networks (Girvan and Newman 2002).

Using percolation theory, Dong et al. (2018) pointed out that in a modular network, a small fraction of nodes that connect nodes of different modules, called 'interconnected nodes', is critical to the robustness of the network. By analyzing the *LCC* size during node removal process by varying the fraction of interconnected nodes (*r*) in the network, they found that *LCC* scale with *r* by a power-law with universal criticality. This result suggests that modular network with higher fraction of interconnected nodes (therefore low modularity *Q* because the fraction of links between nodes in the same modules is lower) will result in a lower *LCC* decrease during node removal and consequently higher network robustness.

Shai et al. (2015) developed both analytical and simulation analyses for evaluating the robustness of random and scale-free model networks with modular structure (Shai et al. 2015). They simulate the attack of interconnected nodes, i.e. nodes that connect to neighbors that are in other modules, and analyze the critical node occupation probability $p_c$, i.e. the fraction of remaining *p* when a large decrease in *LCC* occurs, as a function of the number of modules *m* and the ratio between probabilities for an intra- and inter-module link *α*.

They found that percolation phase transition falls into two regimes depending on the number of modules *m* for a fixed *α*:

- For $m < m^*$ the network presents very high modularity and collapses abruptly under node removal (i.e. $1^{st}$ order phase transition) as a result of the modules becoming disconnected from each another, while their internal structure is almost unaffected.
- In contrast, for $m > m*$, the network presents low modularity and the interconnected nodes play an important role to maintain the whole network connected when nodes are removed. Therefore, the node attack causes lower damage breaking continuously the entire system without sharp *LCC* decrease (i.e. $2^{nd}$ order phase transition). Put another way, $m^*$ represents the threshold above which the network modular structure vanishes and the network returns to behaving as a non-modular network.

In this work, we analyze how the modularity of scale-free model and real-world social networks affects their robustness and the efficacy of the node attack strategies. Using model network, we vary the level of modularity by changing the ratio of intra-modules links over inter-modules links (*κ*). We found that the attack strategy based on node betweenness, which was found to be the most effective strategy to break the *LCC* of real-world networks (Wandelt et al. 2018; Nguyen et al. 2019), is the best strategy to disrupt the *LCC* only when *κ* is higher than a given value $κ_c$, i.e. when the network has high modularity. Below, when network has low modularity *Q*, or even no modular structure, the attack strategy based on node degree is more effective. In addition, the type of the network percolation phase transition when nodes are removed change from a continuous 2nd order (in which *LCC* has no abrupt decrease) to an abrupt 1st order transition (with abrupt *LCC* decrease) when $\kappa$ increases. We also examine the effect of network's density (i.e. the average number of links per node) and the number of modules on network robustness and found that those parameters affect the network robustness, but not the type of the network percolation phase transition (1st or 2nd order) which only depends on $\kappa$. Finally, we study those effects for a variety of real social networks and we found that real social networks with higher modularity *Q* are less robust when subjected to the attack strategy based on nodes betweenness. In other words, the efficacy of the attack strategy based on nodes betweenness is higher for real social networks showing higher modularity *Q*.

**Methods**

A network can be represented as a graph $G = (V, E)$, where $V = \{1,2,...,N\}$ is the set of *N* nodes (vertices), and $E = \{e_{ij} \mid i, j \in V, i \neq j\}$ is the set of *E* links (edges). Networks can be undirected when the links have no specified direction, or directed, in the case links present directionality. Network are unweighted when only the presence-absence of the links is considered, or weighted, in the case some interaction value is associated to the link, i.e. the link weight. Undirected and unweighted networks can be abstracted by an adjacency an *NxN* matrix A where element $a_{i,j}=1$ when there is a link between node *i* and *j* and $a_{i,j}=0$ otherwise. In this paper, only undirected and unweighted networks are considered.

Generation of model scale-free network

Model scale-free networks with size of $N = 10000$ nodes are generated using the well-known Barabási- Albert (BA) model (Barabasi and Albert 1999). The BA model starts from a small clique (a completely connected graph) of $N_0$ nodes. At each successive time step, a new node is added and connected to $M_0$ different existing nodes ($M_0 < N_0$) with the probability of connect an existing node is proportional to its degree (i.e the number of links to the node). The network then has a power-law degree distribution $P(k)=k^{-\gamma}$ with degree exponent $\gamma = 3$ (Barabasi and Albert 1999). We chose the average node degree $<k>$ between 2 and 32.

Our model to generate modular scale-free network

From the BA network we generated modular networks using a new *ad hoc* algorithm by rewiring links as following:

- Each node is assigned randomly to a module $c_i = \{1,2,\ldots,m\}$ where $m$ is the total number of modules. The number of nodes in each module is approximately $N/m$.
- For each link connecting two nodes $i$ and $j$ of different modules $c_i \neq c_j$ (inter-modules links), we will rewire it with a probability $w$ (and keep it without rewiring with probability $1 - w$) by the following procedure:
    - We randomly select one node between the two ending nodes of the link, says $i$, and find another node $l$ within the same module of the node $i$ ($c_l = c_i$). We then detach the inter-modules link between nodes $i$ and $j$ and create a new intra-module link between nodes $i$ and $l$. The node $l$ is selected with a probability proportioned to its degree (node with higher degree in the module $c_i$ has higher probability of being selected)
    - If some nodes are isolated in the network after rewiring, they will be removed. However, we find that only a negligible proportion of nodes can be isolated after the rewiring.

We show in Appendix A that, as long as $N$ is high enough, this rewiring procedure statistically preserve the BA model node degree distribution (scale-free and degree exponent $\gamma = 3$).

Thus, by changing the probability $w$ we can change the ratio $\kappa$ between the number of intra-module links $L_{intra}$ (links that connects two nodes from the same module) and inter-modules links $L_{inter}$ (links that connects two nodes of different modules), thus varying the community structure of the network. The relation between $\kappa$ and $w$ can be derived as following:

- The number of inter-modules links and intra-module links before the rewiring process are $\frac{(m-1)}{m}N\langle k \rangle$ and $\frac{1}{m}N\langle k \rangle$, respectively.
- After the rewiring process, the number of inter-modules links become $(1-w)\frac{(m-1)}{m}N\langle k \rangle$ and the number of intra-modules links become $(\frac{1}{m} + w\frac{(m-1)}{m})N\langle k \rangle$
- The ratio between the number of intra-module links ($L_{intra}$) and inter-modules links ($L_{inter}$) become

$$\kappa = \frac{L_{intra}}{L_{inter}} = \frac{1 + w(m-1)}{(1-w)(m-1)} = \frac{m}{(1-w)(m-1)} - 1$$

which is a monotone function of $w$ when $m > 1$.

- We derive $\alpha$, the ratio between the probability for a given link to be intra-link ($p_{intra}$) over that for a given link to be inter-link ($p_{inter}$) as in (Shai et al 2015) by:

$$\alpha = \frac{p_{intra}}{p_{inter}} \sim (m-1)\frac{L_{intra}}{L_{inter}} = \frac{m}{(1-w)} - (m-1)$$

which is also a monotone function of $w$ when $m > 1$.

The monotone change of $\kappa$ and $\alpha$ as function of $w$ was confirmed with simulation results which are shown in Appendix B.

Thus increasing $w$, we increase the modularity of the network, i.e. increasing $w$ we marked the network community structure. Figure 2 presents example of modular network with $m = 5$ and different value $w$, created from the initial network with $N = 10000$ and the average degree of $<k> = 8$.

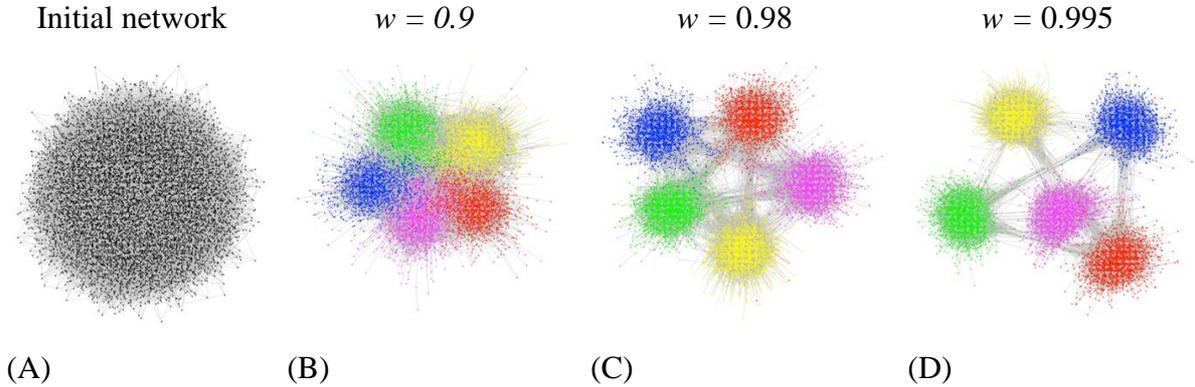

(A)    (B)    (C)    (D)

**Figure 2**: Visualization of the model for generating scale-free modular networks. The modularity of the network increases from A to D. (A) initial non-modular scale-free BA network of size $N = 10000$ with average degree $<k> = 8$. (B)-(D): Illustration of the increasing modularity effect of $w$ on the obtained modular network divided into $m = 5$ modules.

<u>The node attack strategies</u>

The network generated above will be exposed to two node attack (removal) simulation processes (or node attack strategy) where a $p$ proportion of nodes with lowest centrality measures are kept and $q = 1-p$ proportion of highest centrality measure nodes are removed together with their links:

- The first attack strategy removes nodes according to their degree, i.e. the number of links to it, as centrality measure and it is called initial degree (ID) based attack strategy (Albert and Barabasi 2002, Bellingeri et al. 2014, Wandelt et al. 2018).
- The second strategy uses a macro-scale network metric, the node betweenneess, which is the number of times that a node appears in the shortest paths among all nodes pairs in the network (Brandes 2001). This method is commonly used to break real-world networks and is called initial node betweenneess attack strategy (IB) (Bellingeri et al. 2014; Wandelt et al. 2018).

The network robustness measures

To measure the robustness of the network under nodes attack we traced the size of first largest connected component 1$^{st}$ *LCC* and the second 2$^{nd}$ *LCC* as a function of *p*. Further, for each attack simulation, we compute a single value defined as the network robustness (*R*) as done in Bellingeri et al. (2019a). The value of *R* is the average of the normalized sized of the 1$^{st}$ *LCC* (normalized by the initial number of node *N*) along the removal process. *R* can range between two theoretical extremes, $R \simeq 0$ (absolute fragile network) and $R \simeq 1$ (absolute robust network). In addition, we identify the critical value of occupation probability $p_c$ as the largest value of *p* where *LCC* has an abrupt decrease, as shown in Figure 1B. In the case no abrupt decrease was found (Figure 1A), we compute $p_c$ using the "Molloy-Reed" criterion (Callaway et al. 2000; Cohen et al. 2000), which states that the network loses its overall connectivity when each node in the network has less than two links on average. It translates to the mathematic condition of $<k^2>/<k> < 2$, where $<k>$ is the node degree. Thus, the higher are *R* and the lower $p_c$ the more robust is network under node attack. As a consequence, when comparing the efficacy of the node attack strategies, the higher are *R* and lower $p_c$, the lower is the efficacy of the strategy to disrupt the *LCC*. We then denote $p_c^{ID}$ and $p_c^{IB}$ the node occupation probability against ID and IB node attack strategies, respectively; as well as we denote $R_{ID}$ and $R_{IB}$ the network robustness against ID and IB node attack strategies, respectively.

Real-world social networks dataset

In addition to model networks, we analyze 12 real-world social networks, in which 8 are networks of Facebook's pages where nodes represent pages of different topics - TV Shows, Politician, Government, Public Figures, Athletes, Company, New sites and Artist - and links are mutual likes between them. The Facebook's pages data is collected from

https://snap.stanford.edu, prepared by (Rozemberczki et al. 2019). Beside, we use two financial networks where nodes represent the US SP500 stocks and links are calculated from the correlation matrix using threshold method (see Nguyen et al. 2019 for more detail); the co-authorship network of scientists working on network theory and experiment (NetScience) where nodes represent authors and link's weight represents the number of common papers (Newman 2003; Boccaletti et al. 2006); and the Email network of people in a large European Research Institution (Email) where nodes represent researchers and links indicate that at least one email was sent between two researchers (Leskovec et al. 2007; Yin et al. 2017).

Table 1 summarizes the following statistics of the real-world social networks:

- Node degree: is the number of links to the node (Boccaletti et al. 2006). The degree of node *i* is given by:

$$k_i = \sum_{j \neq i \in N} a_{ij}$$

where $a_{ij}=1$ in the case there is a link connecting nodes *i* and *j* and is 0 otherwise; the term *N* means the sum is over all nodes in the network.

- Modularity: The modularity indicator *Q* calculates how modular is a given division of a network into subnetworks (modules or communities):

$$Q = \frac{1}{2L} \sum_{i,j} (a_{ij} - \frac{k_i k_j}{2L}) \delta(c_i, c_j)$$

where *L* is the number of links, $a_{ij}$ is the element of the A adjacency matrix in row *i* and column *j*, $k_i$ is the degree of *i*, $k_j$ is the degree of *j*, $c_i$ is the module (or community) of *i*, $c_j$ that of *j*, the sum goes over all *i* and *j* pairs of nodes, and $\delta(x, y)$ is 1 if x = y and 0 otherwise (Clauset et al. 2004).

- LCC: the largest connected component (also called 'giant cluster') represents the maximum number of connected nodes in the network (Boccaletti et al. 2006; Bellingeri et al. 2020). Considering all the network clusters, i.e. the sub-networks of connected nodes, the LCC can be defined:

$$LCC = \max_j (S_j)$$

where $S_j$ is the size (number of nodes) of the $j$-th cluster.

- Diameter: the diameter of the network ($D$) is the longest shortest path length of all pairs of nodes in the network, also called the longest geodesic (Newman 2013)
- Transitivity: the transitivity ($C$) is based on triplets of nodes. A triplet is three nodes that are connected by either two (open triplet) or three (closed triplet) undirected edges. The transitivity is the number of closed triplets (or 3 x triangles) over the total number of triplets (both open and closed). In formula:

$$C = \frac{\lambda_{closed}}{\lambda_{total}}$$

where $\lambda_{closed}$ is the number of closed triples and $\lambda_{total}$ is the number of all possible triples in the network. Transitivity represents the overall probability for the network to have adjacent nodes interconnected, thus making more tightly connected modules (Newman et al. 2002)

- Link Density: the link density (Density) is number of links divided by the total number of possible links (Boccaletti et al. 2006).

| Networks | $N$ | $L$ | LCC | LCC(%) | <k> | D | C | Density | Q |
|---|---|---|---|---|---|---|---|---|---|
| TV Shows | 3,892 | 17,262 | 3,892 | 100% | 4.4 | 20.0 | 0.443 | 0.00228 | 0.830 |
| Politician | 5,908 | 41,729 | 5,908 | 100% | 7.1 | 14.0 | 0.429 | 0.00239 | 0.815 |
| Government | 7,057 | 89,455 | 7,057 | 100% | 12.7 | 10.0 | 0.433 | 0.00358 | 0.614 |
| Public Figures | 11,565 | 67,114 | 11,565 | 100% | 5.8 | 15.0 | 0.215 | 0.00100 | 0.645 |
| Athletes | 13,866 | 86,858 | 13,866 | 100% | 6.3 | 11.0 | 0.303 | 0.00090 | 0.637 |
| Company | 14,113 | 52,310 | 14,113 | 100% | 3.7 | 15.0 | 0.287 | 0.00053 | 0.656 |
| New sites | 27917 | 206,259 | 27,917 | 100% | 16.2 | 15.0 | 0.138 | 0.00052 | 0.529 |
| Artist | 50,515 | 819,306 | 50,515 | 100% | 7.4 | 11.0 | 0.295 | 0.00064 | 0.457 |
| SP500_1 | 315 | 8,706 | 315 | 100% | 27.6 | 6.0 | 0.511 | 0.08802 | 0.253 |
| SP500_2 | 371 | 10,636 | 369 | 99% | 28.7 | 6.0 | 0.718 | 0.07748 | 0.373 |
| NetScience | 1,589 | 2,742 | 379 | 24% | 1.7 | 17.0 | 0.878 | 0.00109 | 0.954 |
| Email_EU | 1,005 | 16,064 | 986 | 98% | 16.0 | 7.0 | 0.450 | 0.01592 | 0.341 |

**Table 1**: Structural statistics of the real-world social networks: nodes (*N*), links (*L*), size of the *LCC*, size of the *LCC* as % with respect the total number of network nodes, average node degree *<k>*, diameter, transitivity, the edge density, modularity *Q*.

**Results**

Robustness of non-modular scale-free network

We simulate scale-free network of size $N = 10000$ nodes and average degree $<k> = 4$ and run attack simulation using ID and IB strategies. The 1$^{st}$ *LCC* decreases continuously under both strategies and the network is completely broken down (i.e. the *LCC* shrinks to *quasi* zero) at a critical occupation probability $p_c$ (0.62 and 0.56 for ID and IB, respectively), as seen in Figure 3A. At this transition, we also found that the 2$^{nd}$ *LCC* has its maximum value as shown in Figure 3B. Such phase transitions are called 'continuous phase transitions' or 'second-order phase transitions' and denote robust network (see Mnyukh 2013). Interestingly, while overall findings showed that nodes attack strategies based on betweenness centrality are highly efficient for most real-world networks (Bellingeri et al. 2014; Iyer et al, 2013; Nguyen and Nguyen 2018; Wandelt et al. 2018), our results shown different conclusion. For scale-free network without modular structures (lower value of parameter *w*), betweenness-based strategy IB does not perform better than the degree-based strategy ID. For scale-free network with significant modular structures (higher value of parameter *w*), betweenness-based strategy IB clearly performs better than the degree-based strategy ID. It is therefore arguable that the presence of modular structure in networks is an important factor enhancing the efficacy of betweenneess-based attack strategy for breaking the 1$^{st}$ *LCC*, as shown in the next sub-section.

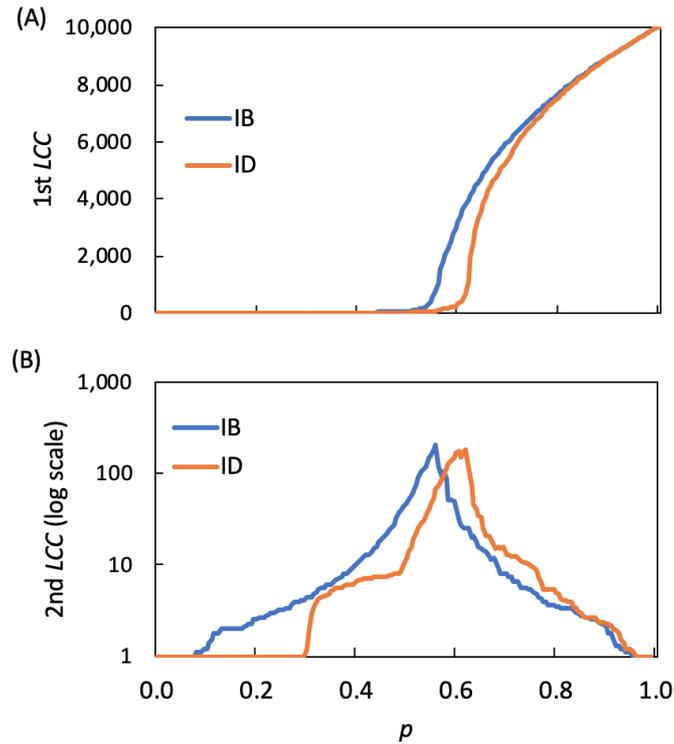

**Figure 3**: Simulation result for the non-modular network with $N = 10000$ nodes and average degree $<k> = 4$: A) Size of the first largest connected component (1st *LCC*) and B) the second largest connected component (2nd *LCC*) as a function of the occupation probability $p$.

**Robustness of modular scale-free network**

We first present the robustness of the network of different modularity by varying the re-wiring ratio $w$, then we discuss the robustness of the network with different node average degree $<k>$ and number of modules $m$.

Robustness as a function of the modularity

We simulate scale-free network of size $N = 10000$ nodes with $m = 5$ modules and average degree $<k> = 4$, then applying the rewiring method with increasing $w$. At first when $w$ is small (and the network presents low modularity), we found that the network is resilient and the $p_c$ remains approximately equivalent value as with original non-modular network for both ID and IB attack strategies (Figure 4). At this degree of modularity, the network is still homogenous enough with a high number of inter-modules links. In consequence, when the attack strategies remove nodes the 1st *LCC* continuously become smaller but still hold the connection between modules, denoting higher network robustness against node attack.

Only when $w$ is higher than 0.95 the network become fragile and the 1st *LCC* abruptly decreases at some value of $p$, as seen in Figure 4A and B when the network is attacked by the IB and ID strategies, respectively. This value of $w = 0.95$ corresponds to the ratio between the number of intra-module links and inter-modules links $\kappa_c = 23.8$ (for $m = 5$). At this point the connection between modules in network is sparse enough and the removal of critical nodes may break down the global connectivity among modules even though the modules are still relatively well connected. We call $p_c$ the largest value of $p$ with an abrupt decrease of the 1st *LCC*, as proposed by (Shai et al. 2015), and show its relationship with $w$ in Figure 5A. This transition corresponds to a first-order phase transition where local structures are separated from the 1st *LCC* (denoting lower network robustness). As a result, the size of the 2nd *LCC* abruptly increase at $p_c$ and gradually decrease afterward (see Figure 4C and D).

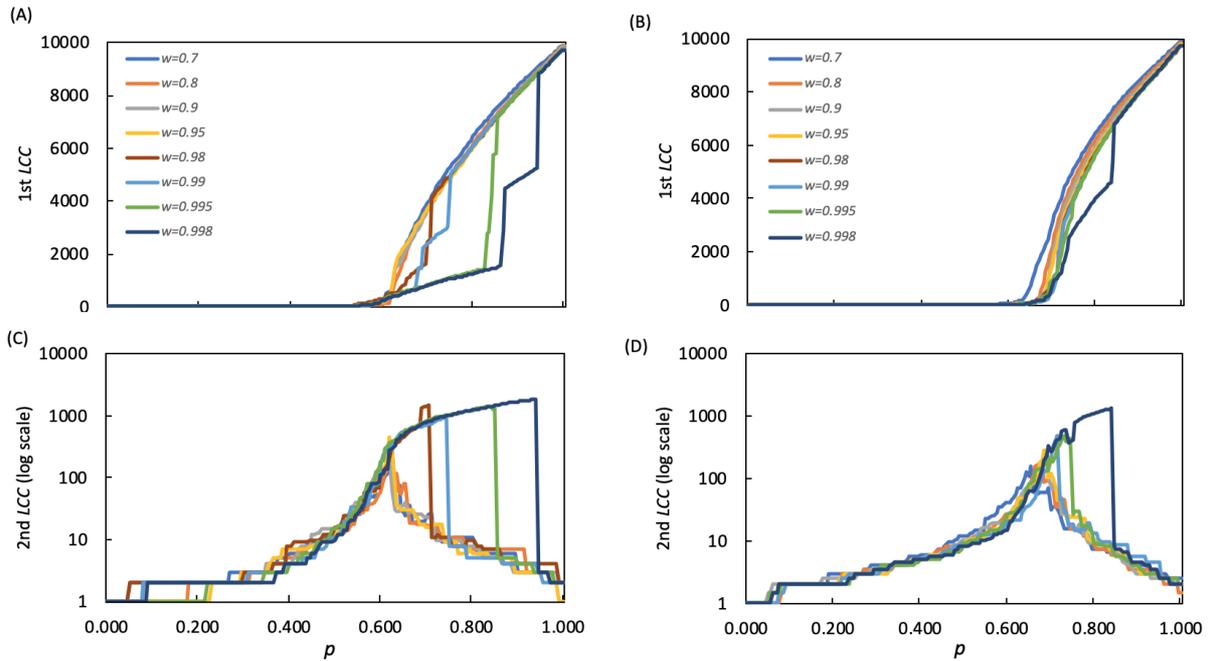

**Figure 4**: Simulation result for modular networks generated from BA network with $N = 10000$, $m = 5$ and $<k> = 4$ with different value of rewiring probability $w$: Size of the first largest connected cluster (1st *LCC*) and the second largest connected cluster (2nd *LCC*) as a function of the occupation probability $p$ when attacked by IB (A and C, respectively) and ID (B and D, respectively).

As shown in Figure 5A, the $p_c^{IB}$ increases faster than the $p_c^{ID}$ and when $w$ is higher than 0.98, $p_c^{IB}$ become higher than $p_c^{ID}$ showing that the network becomes more vulnerable to the IB strategy than the ID strategy. Similar behavior is confirmed with the robustness measure $R$ (Figure 5B).

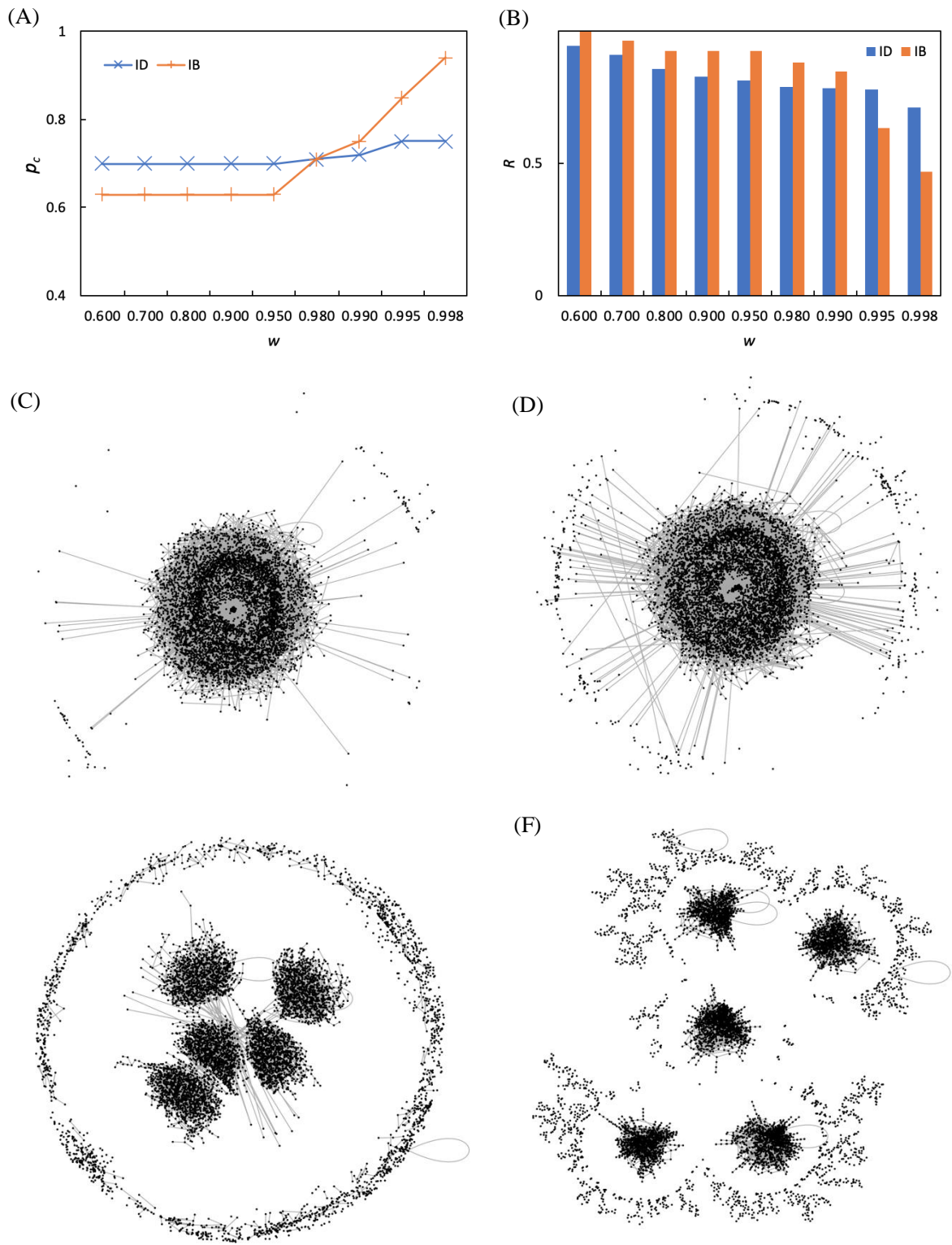

**Figure 5**: Simulation result for modular networks generated from BA network with $N = 10000$, $m = 5$ and $<k> = 4$: (a) The critical occupation probability $p_c$ and (b) the single value network robustness (R) as a function of the re-wiring ratio $w$. The trend is clear, when w is higher than 0.98, $p_c^{IB}$ becomes higher than $p_c^{ID}$ and $R_{IB}$ becomes smaller than $R_{ID}$, showing that the network becomes more vulnerable to the betweenness-based strategy than the degree-based strategy. For illustration we present the the images of two modular networks subjected to node removal: with $w = 0.6$ and attacked by the ID (c) and IB (d) strategies; with $w = 0.998$ and attacked by the ID (e) and IB (f) strategies; In all simulation the occupation probability $p$ is 0.6.

Robustness as a function of network density

Next we examine the effect of the link density (i.e. the average number of links per node) by simulating scale-free network of size $N = 10000$ nodes, number of modules $m = 5$ with varying average node degree $<k>$ from 2 to 16 attacked by the betweenness-based strategy (IB). We found that $p_c$ decreases as the link density increase (Figure 6) - the networks become more robust when nodes have more links. However, the transitions type only depends on the rewiring ratio $w$ and is relatively stable with respect to the average degree $<k>$ change (Figure 6 and 7). In other words, the ratio of probability of inter-modules links over the probability of intra-module links $\alpha$ (which is a function of $w$) is the critical factor to determine the type of the phase transition.

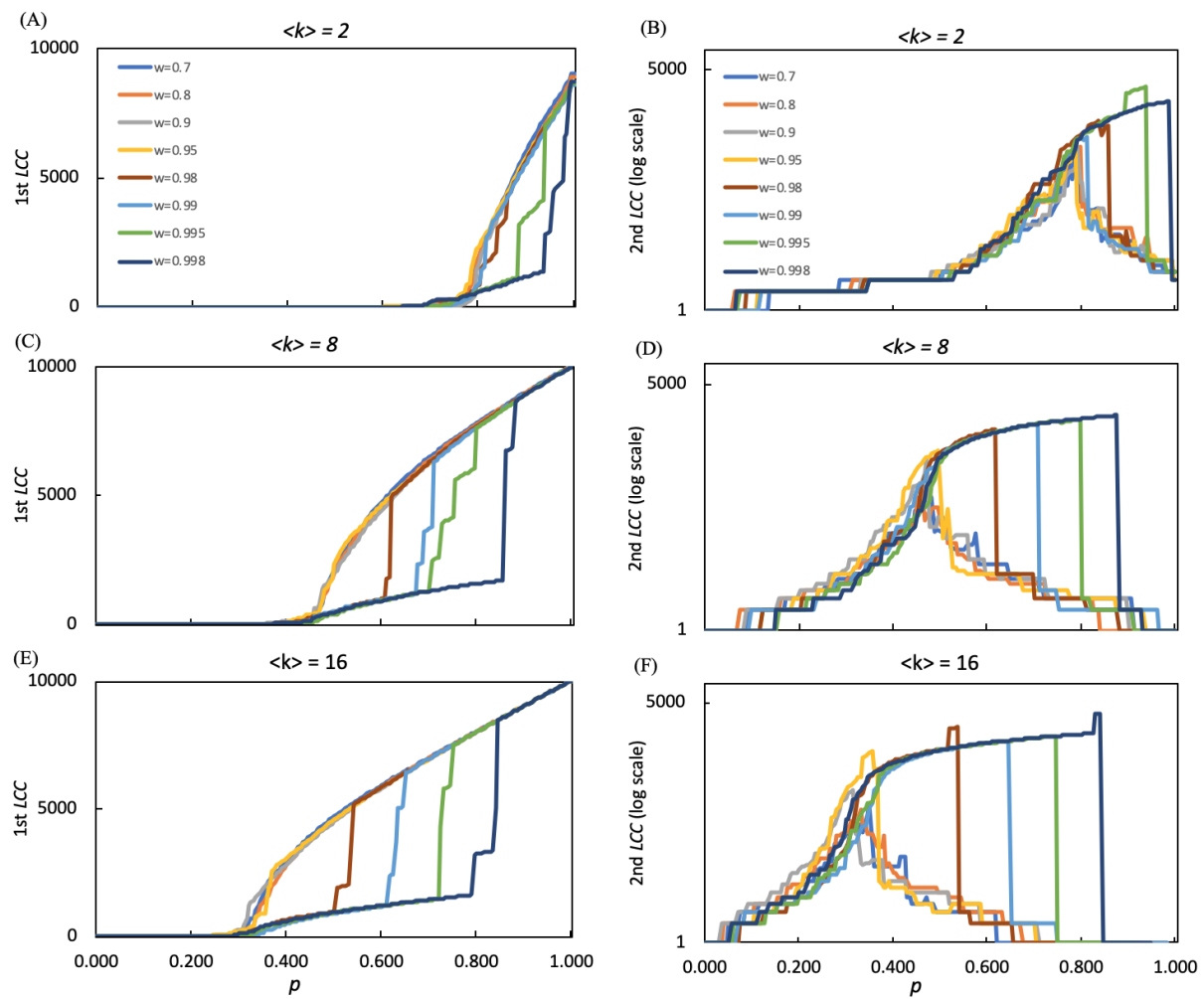

**Figure 6**: Simulation result by IB attack strategy for the modular networks generated from BA network with $N$ = 10000, $m$ = 5 with different value of rewiring probability $w$ and for different node average degree $<k>$. Size of the 1$^{st}$ *LCC* (A, C, and E) and the 2$^{nd}$ *LCC* (B, D and F) as a function of the occupation probability $p$ when attacked by IB for various average node degree $<k>$

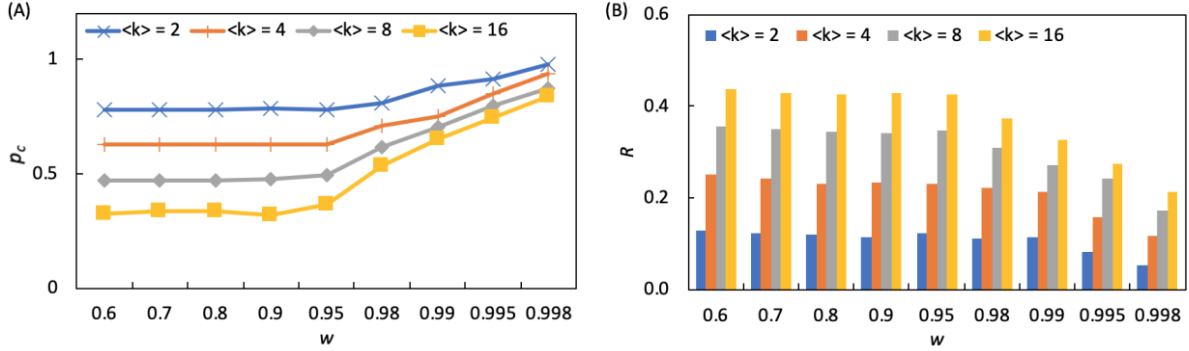

**Figure 7**: Robustness measures of simulation result (IB) for the modular networks generated from BA network with $N$ = 10000, $m$ = 5 with different value of rewiring probability $w$ and for different node average degree $<k>$: (A The critical occupation probability $p_c$ and (B) the single value network robustness ($R$) as a function of the re-wiring ratio $w$ for different node average degree $<k>$.

By number of modules:

Here, we generated scale-free network of size $N$ = 10000 nodes and average degree $<k>$ = 4 with number of modules $m$ varying from 2 to 20. We run node attack simulation by the betweenness strategy (IB) only founding that $p_c$ sharply decreases below $w$=0.98 regardless of the number of modules in the network (Figure 8). In other word, the network robustness (lower $p_c$) sharply increases below $w$=0.98, that is when the network present lower modularity, thus reaching the maximum robustness for *non*-modular network. For this reason, the transition type is insensitive to the number of modules $m$, similar to the effect of network density. Moreover, we found that the critical occupation probability $p_c$ slightly increases with $m$, suggesting that the model networks become slightly more fragile when they have more modules (Figure 8).

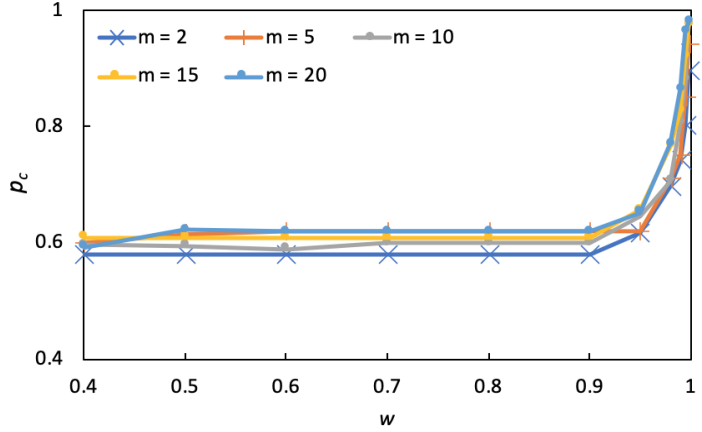

**Figure 8**: The critical occupation probability $p_c$ as a function of the re-wiring parameter $w$ for different number of modules $m$ in the network. The result is obtained with IB attack strategy for the modular networks generated from BA network with $N = 10000$ and $<k> = 4$.

Robustness and structural properties in model and real-world networks:

To verify the relationship between modularity, node degree, and the efficacy of the attack strategies we show above for model networks we investigate for such relationship also for 12 real-world social networks (Table 1). We fit linear models of the robustness $R_{IB}$ against the modularity $Q$ and the average node degree $<k>$. In Figure 9A we show the linear model of the $R_{IB}$ with respect to the modularity $Q$ for our model modular network generated with different $w$ from a BA network of $N = 10000$, $<k> = 4$ and $m = 5$. We find a significant trend as $R_{IB}$ decreases when $Q$ increases ($p$-value= 0.01). Very interesting, in our real social networks dataset, we find a similar $R_{IB}$ decrease with modularity $Q$ ($p$-value< 0.001). Remarkably, we find an abrupt transition in model network, i.e. when $Q$ is very high (about 0.8), the network's robustness drastically decreased (Figure 9A). We do not find this effect in real-world social networks (Figure 9C). The absence of an abrupt robustness $R_{IB}$ decrease in real-world social networks can be due to the fact that real networks may vary in link density and other structural properties (for example, transitivity, assortativity, number of modules, etc..) that may affect the network response to IB node attack. For this reason, the variability in real-world social networks structure, with many structural factors affecting the network robustness, may prevent the abrupt $R_{IB}$ decrease as a function of the modularity $Q$ that we observe in model networks.

Further, we find a clear $R_{IB}$ increase by increasing the average node degree in our model networks ($p$-value< 0.001, Figure 9B). This is in agreement with past analyses showing how the network robustness to node removal increase with the linkage density, i.e. the higher the number of links per node, the slower is the network fragmentation under node removal (Albert and Barabasi 2002; Iyer et al. 2013). Differently, we do not find a significant relationship between $R_{IB}$ and $<k>$ in our real social networks dataset (*p*-value=0.16, Figure 9D). Even in this case, the variability in real-world networks structure, with many structural factors affecting their robustness, may hide the emergence of a clear relationship between the linkage density measured by the average node degree $<k>$ and the robustness of the network ($R_{IB}$) against IB node attack.

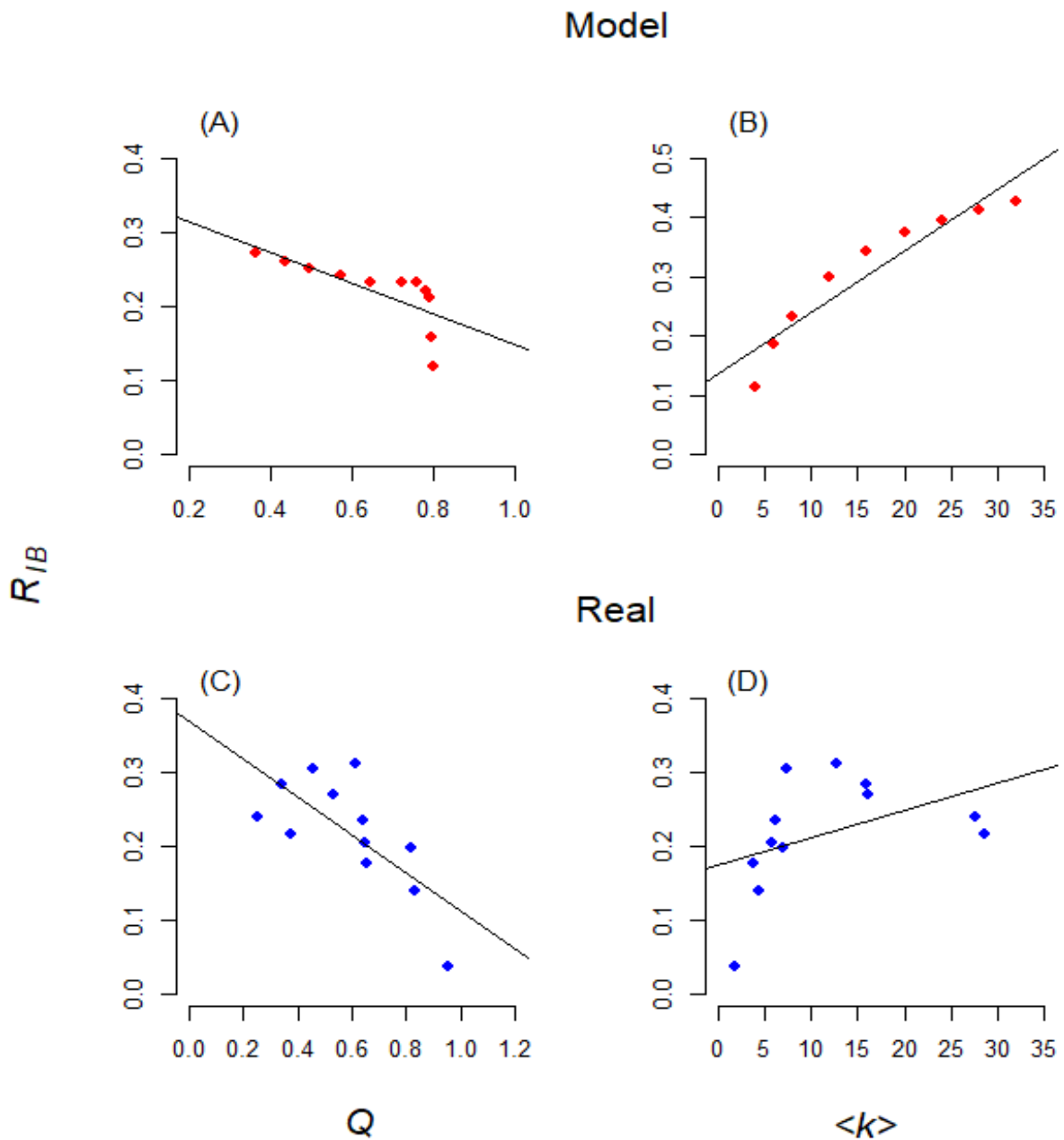

**Figure 9**: Linear models of the robustness $R_{IB}$ as a function of the modularity indicator $Q$ ($R_{IB}=\alpha \cdot Q + \beta$, left column) and the average node degree $<k>$ ($R_{IB}= \alpha \cdot <k> + \beta$, right column). Top row: model networks (panels A, B, red points); Bottom row: real-world social networks (panels C, D, blue points). (A) Model networks are generated with $N = 10000$, $<k> = 4$, $m = 5$ and with increasing modularity by varying the parameter $w$. (B) Model networks are generated with $N = 10000$, $m = 5$, $w = 0.9$ and varying $<k>$ in the interval (4,32). Statistical outcomes of the linear model parameters (intercept $\alpha$ and slope $\beta$): (A) $R_{IB}$ vs $Q$: intercept $\alpha =-0.2$, slope $\beta =0.35$, p-value= 0.01; (B) $R_{IB}$ vs $<k>$: $\alpha =0.01$, $\beta =0.14$, p-value<0.001; (C) $R_{IB}$ vs $Q$: $\alpha =-0.26$, $\beta =0.37$, p-value= 0.001; (D) $R_{IB}$ vs $<k>$: $\alpha =0.004$, $\beta=0.18$, p-value= 0.16.

**Discussion and Conclusion**

In this work we study the robustness of model scale-free networks and real-world social networks with different modularity. The model networks are generated from BA model with a novel method for tuning their modular structure. Using Monte-Carlo simulation we simulate two node attack strategies, IB and ID based on node's betweenness and degree, respectively. With both attack strategies, we found two types of percolation phase transitions take place. The 1st order abrupt phase transition happens when the model network has high modularity, representing by $\kappa > \kappa_c$ with $\kappa_c \sim 23.8$ for both IB and ID attack strategies. Also at and above this critical point, the model network is more fragile under betweenness-based strategy attack: $R_{IB} < R_{ID}$, as found in many real-world complex networks. When $\kappa < \kappa_c$ or when the model network has no modular structure, the network experiences a continuous 2nd order phase transition under both nodes attack strategies. Interestingly, under this regime, the network is more robust against the betweenness-based attack strategy IB than the degree-based attack strategy ID, contrary to most of results on real-world networks. In addition, our work showed that the ratio $\kappa$ is the main factor for the type of phase transition: small $\kappa$ corresponds to 2nd order continuous phase transition while high $\kappa$ corresponds to 1st order abrupt phase transition. Further, we investigate how the modularity affect the robustness of the system against node removal in 12 real-world social networks and find a similar $R_{IB}$ decrease with modularity $Q$ (p-value< 0.001) that we observe in model networks varying the modularity. This result indicate how network with higher modularity (i.e. with higher community structure) may be more fragile to betwenness based node attack. At the same, this result show how the betwenness based node attack (IB) is highly effective when attacking network with marked community structure (higher modularity $Q$). Differently, in the case the network shows very low modularity (or no modularity), the degree-based node attack ID may perform better than IB.

This result helps to understand the role of modularity and community structure for the robustness of networks, to select most effective node removal in networks, and may shed light to the design of robust networks.

**Acknowledgement**

This work is supported by the Vietnam's Ministry of Science and Technology (MOST) under the Vietnam-Italy scientific and technological cooperation program for the period of 2021-2023. Many thanks to Dr. Vu-Lan Nguyen for useful comments on the paper.

**APPENDIX**

**A. Prove that the rewired network statistically preserves the original node degree distribution**

Given a node with degree $k$, the proportion of inter-modules links and intra-module links of this node before the rewiring process are approximated by $\frac{(m-1)}{m}k$ and $\frac{1}{m}k$, respectively, where $m$ is the number of modules. A proportion $w$ of its inter-modules links will be rewired, thus the expected number of links that this node loses is:

$$w\frac{(m-1)}{m}k$$

Similarly, this node can also be selected when links from nodes of the same modules are rewired. We compute the expected number of rewired links that this node can acquire as following:

- The total of rewired links in the network is $w\frac{(m-1)}{m}N<k>$
- The total of rewired links that will be connected to nodes within the module of the node is: $w\frac{(m-1)}{m}N<k>/m$
- The probability that the node is selected is proportioned to the ratio of its degree to the total degree of all nodes in the module (according to our method) and is: $k/(N<k>/m)$
- The expected number of rewired links that this node can be selected is therefore equal to: $w\frac{(m-1)}{m}N<k>/m \times k/(N<k>/m) = w\frac{(m-1)}{m}k$ which is exactly equal to the expected number of links that this node loses. In consequence, the expected number of links of each node after rewiring process is equal to their initial degree, and the network's degree distribution remain unchanged.

**B. Graph of $\kappa$ and $\alpha$ as function of rewiring probability $w$ and number of modules $m$**

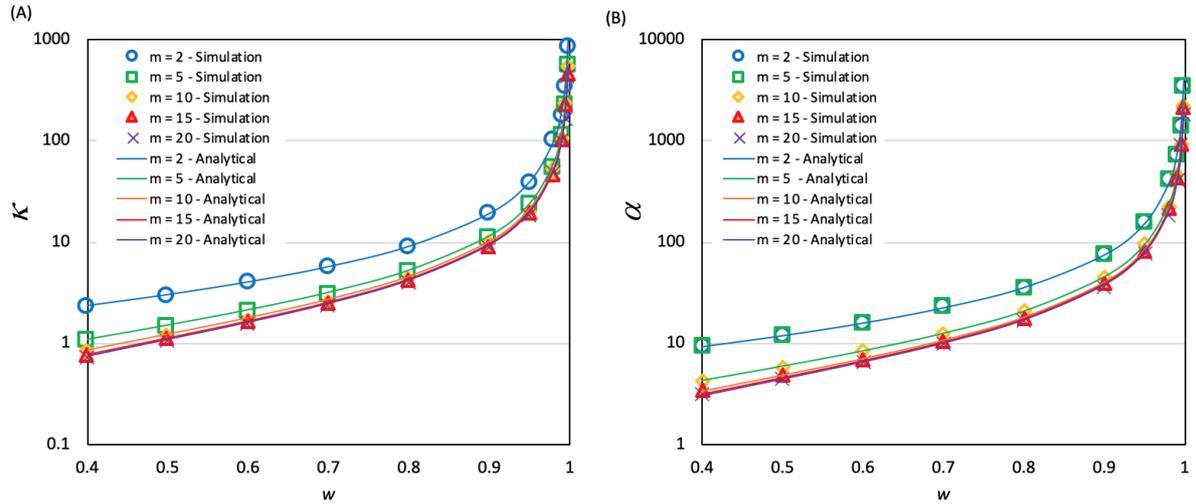

Comparison of analytical and simulation results for modular scale-free network for A) $\kappa$ as a function of $w$ and $m$ to show and B) $\alpha$ as a function of $w$ and $m$. Both measures show the goodness of mathematical derivation in the Method section.